\documentclass[11pt]{article}%[amssymb,11pt]
\usepackage[pdftex,colorlinks=true,urlcolor=black,filecolor=black,linkcolor=black,
            pdftitle={Hybrid imploding scalar and ads spacetime.},
            pdfauthor={Mark D. Roberts},
            pdfsubject={Anti-deSitter},
            pdfkeywords={imploding scalar},
            pagebackref,pdfpagemode=None,bookmarksopen=true]{hyperref}
\usepackage{graphicx}
\newcommand{\bc}{\begin{center}}
\newcommand{\ec}{\end{center}}
\newcommand{\be}{\begin{equation}}
\newcommand{\ee}{\end{equation}}
\newcommand{\ber}{\begin{eqnarray}}
\newcommand{\ear}{\end{eqnarray}}
\newcommand{\ch}{\chi}

\newcommand{\fr}{\frac}

\newcommand{\lb}{\label}

\newcommand{\n}{\nonumber\\}

\newcommand{\st}{\stackrel}

\begin{document}
%%%%%%%%%%%%%%%%%%%%%%%%%%%%%%%%%%%%%%%%%%%%%%%%
%\vspace{-1.1in}
\title{Hybrid imploding scalar and ads spacetime.}
%\title{Hybrid imploding scalar and ads spacetime.}
\author{
\href{http://www.violinist.com/directory/bio.cfm?member=robemark}
{Mark D. Roberts},\\
}
\date{$6^{th}$ of May 2019}
\maketitle
\vspace{1cm}
\begin{abstract}%%%%%%%%%%%%%%%%%%%%%%%%%%%%%%%%%%%%%%%%%%%%%%%%%%%%%%%%%%%%%%%%%%%%%%%%%%%%%
A solution to the massless scalar cosmological constant field equations is presented.
The solution has imploding scalar and parts of anti-deSitter or deSitter spacetime
as limiting cases.   Some of the solutions properties are discussed however not much
can be said because of the contrasting properties of imploding scalar and deSitter
spacetimes.   What can be said is that the self-similar homeothetic Killing vector
is not qualitatively changed by the cosmological constant,  in other words self-similarity
and the presence of a cosmological constant are separate properties.
\end{abstract}%%%%%%%%%%%%%%%%%%%%%%%%%%%%%%%%%%%%%%%%%%%%%%%%%%%%%%%%%%%%%%%%%%%%%%%%%%%%%%%
%\begin{center}\href{http://arXiv.org/abs/1412.8470}
%                                    {\tt 1412.8470}\end{center}
\vspace{1cm}
\tableofcontents
\newpage
\section{Introduction}\label{intro}%%%%%%%%%%%%%%%%%%%%%%%%%%%%%%%%%%%%%%%%%%%%%%%%%%%%%%%%%%%
In special relativity the d'Albertian has spherically symmetric solution $\phi=f(v)/r$,
where $f$ is a suitably differentiable function of the null coordinate $v$,
this allows creation of a singularity of the field $\phi$ out of nothing
at the origin of the coordinates $r=0$.
In general relativity this generalizes to solutions of the scalar-Einstein equation,
one can have solutions with $\phi$ of the same form but then one needs a compensating
null radiation field,  if the null radiation field is taken to vanish then one ends up
with a simple scalar-Einstein solution \cite{mdr86,mdr89} which has a homeothetic Killing
vector \cite{brady};  again one has a scalar field singularity at the origin of the coordinates
and there is also a singularity of the spacetime curvature.
Similar solutions presumable exist in most scalar-tensor theories,  for example variable
Newtonian 'constant' $G$ theories of Clifton et al \cite{clifton} and those which show that
Palatini variation leads to more complex connections than the Christoffel connection \cite{mdrfd}.
Applications of the solution include the following four.
{\it Firstly} to cosmic censorship:  a variant of Israel's theorem shows that in most cases 
static scalar-Einstein spacetimes do not have event horizons \cite{chase} and the existence of 
the solution shows that this is also the case in one particular instance in dynamic spacetimes,
see also \cite{GJ}.   Astrophysical observations \cite{NM}
and \cite{AKL} are thought by some to indicate if event horizons exist;  however what will be 
observed is disk size and this could be infinitesimally larger than $r=2m$ and there be no way 
of disinguishing cases,  or the spacetime could be non-vacuum. 
Models have been built which already assume event horizons and then investigate what happens to
marginally trapped tubes and dynamical horizons Booth et al \cite{BBGB}.
There are one parameter models in which the one parameter is related to both the cosmological
constant and the scalar field which can have either event horizons or not Baier et al \cite{BNS}.
There could be examples in electromagnetic theory similar to scalar field models see
Crisford and Santos \cite{CS}.
There are other ways of looking at cosmic censorship rather than through examples.
The most common is to consider it as an intial value problem for the scalar-Einstein equations,
apparently cosmic censorship works Christodoulou \cite{bi:chr} and all the violating
examples are in some sense of measure zero.
Perturbations of collapse have been studied by Frolov \cite{frolov}.
Analog models \cite{torres} are more concerned with rotaional properties at long distances
rather than singularity structure at short distance.
{\it Secondly} it is the critical case between explosion and implosion in numerical models of 
gravitational collapse \cite{choptuik,gundlach};   what the present solutions show is that the
presence of a cosmological constant does not change things,  as here only existence of
solutions is looked at and not uniqueness there might be solutions with more intricate behaviour,
{\it Thirdly} the induced scalar field in quantum field theories on curved spacetimes can be
equated to the scalar field of the exact solution,  this would constitute a new appoach
however there are many technical problems concerning
whether objects such as the van Vlech determinate converge fast enough.
{\it Fourthly} to canonical quantum gravity,  see \cite{mdrqs}.
So far unsuccessful effort has gone into attempting to add additional parameters
such as mass,  rotation and electric charge to the exact solution.
Recent observations of dark energy are most easily described by a cosmological constant.
Here a solution is given which adds a cosmological constant to the original imploding solution:
there is no explicit interaction between the scalar field and the cosmological constant in the
sense that that the cosmological constant does not appear in the scalar field.
\section{Two parameter solution}\label{two}%%%%%%%%%%%%%%%%%%%%%%%%%%%%%%%%%%%%%%%%%%%%%%%%%%%%%
The field equations are taken to be
\begin{equation}
R_{a b}=\Lambda g_{ab}+2\psi_a\psi_b,
\label{fldeq}
\end{equation}
where $\Lambda$ is the cosmological constant and $\psi$ a scalar field.
In double null coordinates the solution is
\begin{eqnarray}
&ds^2=\Omega^2\left[-dudv+r_+r_-d\Sigma_2^2\right],~~~~~~
d\Sigma_2^2\equiv d\theta^2+\sin(\theta)^2d\phi^2,\\
&2r_{\pm}\equiv(1\pm2\sigma)v-u,~~~
\Omega_{\pm}=\left(1\pm\frac{\Lambda uv}{12}\right)^{-1},~~~
\psi=\frac{1}{2}\ln\left(\frac{r_-}{r_+}\right),\nonumber
\label{solution}
\end{eqnarray}
%u, v the wrond way around in r_\pm in v1,v2,v3
where $\sigma$ is the scalar charge and $r=r_+,~\Omega=\Omega_+$.
%not \Omega=\Omega_-$ as in v1,v2,v3

The solution has a conformal Killing potential
\begin{equation}
K=cuv\Omega,~~~
K\cdot K=-2cuv\Omega^2,~~~
K_{a;b}=-2c\frac{\Omega}{\Omega_+}g_{ab},
\label{Kdef}
\end{equation}
where $c$ is a constant;  note that $K$ can be null,  spacelike or timelike.
For $\Lambda\ne0$,  $K_a=12c\Omega_a/\Lambda$ and $\Omega_a$ can be taken to be the conformal
Killing vector.   The first derivative of the scalar field $\psi$ obeys
\begin{equation}
\psi\cdot\psi=\frac{\sigma^2 uv}{\Omega^2r_+^2r_-^2},~~~
K\cdot\psi=0,
\label{psiprops}
\end{equation}
with which using the field equations (\ref{fldeq}) invariants constructed from the Ricci tensor
can be constructed;
also the Weyl scalars are given by $3\Psi_2=\psi\cdot\psi$.
A null tetrad is
\begin{equation}
n_a=-\delta^u_a,~~~
l_a=\frac{1}{2}\Omega^2\delta^v_a,~~~
m_a=\Omega\sqrt{r_+r_-}\left(\delta^\theta_a+\sin(\theta)\delta^\phi_a\right).
\label{nptetrad}
\end{equation}
The null surface projection,  second fundamental form,  normal fundamental form,
surface gravity and surface stress are defined by
\begin{eqnarray}
&q_{ab}\equiv g_{ab}-2l_{(a}n_{b)},~~~
\chi^{(l)}_{a}\equiv l_{d;c}q^c_aq^c_b,~~~
\eta^{(l)}_a\equiv l_{d;c}q^c_an^d,~~~
\omega^{(l)}\equiv-l_dn^cn^d_{.;c},\n
&\tau^{(l)}_{ab}\equiv-\chi^{(l)c}_{~~~c}n_an_b-2\eta^{(l)}_{(a;b)}-\omega^{(l)}q_{ab},
\label{defsurface}
\end{eqnarray}
%\begin{equation}
%q_{ab}\equiv g_{ab}-2l_{(a}n_{b)},
%\chi^{(l)}_{a}\equiv l_{d;c}q^c_aq^c_b,
%\eta^{(l)}_a\equiv l_{d;c}q^c_an^d,
%\omega^{(l)}\equiv-l_dn^cn^d_{.;c},
%\tau^{(l)}_{ab}\equiv-\chi^{(l)c}_{~~~c}n_an_b-2\eta^{(l)}_{(a;b)}-\omega^{(l)}q_{ab},
%\label{defsurface}
%\end{equation}
respectively.  In the present case just the first term contributes to the surface stress giving
\begin{equation}
\tau^{(l)}_{ab}=-\frac{\Omega}{2r_+r_-}\left(v-u+\frac{\Lambda}{12}v^2(u-(1-4\sigma^2)v)\right)\delta^{uu}_{ab}.
\label{surfacestress}
\end{equation}
At $v=0$ the scalar field $\psi$ vanishes and the metric is the same as flat spacetime,
the easiest way to see this is to note that $g_{\theta\theta}=r^2\exp(2\psi)$
then when substituting $\psi=0$ this gives flat spacetime.
Now
\begin{equation}
\tau^{(l)}_{ab}|_{v=0}=\frac{\Omega u}{2r_+r_-}\delta^{uu}_{ab},
\label{restrictedss}
\end{equation}
and as this is independent of both $\sigma$ and $\Lambda$ the junction is well-defined,
the contary has sometimes been claimed \cite{szabodos}.
\section{Three parameter solution}\label{tps}%%%%%%%%%%%%%%%%%%%%%%%%%%%%%%%%%%%%%%%%%%%%%%%%%
In double null $u,v$ coordinates the three parameter solution is
\begin{eqnarray}\label{three}
&ds^2=\Omega^2\left[-dudv+r_+r_-d\Sigma_2^2\right],~~~~~~
d\Sigma_2^2\equiv d\theta^2+\sin(\theta)^2d\phi^2,\\
&2r_{\pm}\equiv(1\pm2\sigma)u-v\pm2\beta,~~~
\Omega_{\pm}=\left(1\pm\frac{\Lambda uv}{12}\right)^{-1},~~~
\psi=\frac{1}{2}\ln\left(\frac{r_-}{r_+}\right),\nonumber
\end{eqnarray}
where $\sigma$ is the dynamical scalar charge,  $\beta$ is the static scalar charge
and $\Lambda$ is the cosmological constant.
For $\sigma=0$ it is static.
For $\beta=0$ it is the solution \cite{mdrcs}.
Many of the simple relationships of \cite{mdrcs} between the various geometric objects do
not survive the generalization to three parameters and become complicated and messy so that
they are not reproduced here.
\section{Hyperbolic dual}\label{hd}%%%%%%%%%%%%%%%%%%%%%%%%%%%%%%%%%%%%%%%%%%%%%%%%%%%%%%%%%%%
Restricting attention in (\ref{three}) to the $\beta=0$ case let $r=r_+,~\Omega=\Omega_-$.
There is a Killing potential is $K=uv$ which when differentiated gives a homeothetic
Killing vector.

To transfer to single null coordinates use $r=r_+$ then
\begin{equation}\label{origsn}
ds^2=-(1+2\sigma)dv^2+2dvdr+r(r-2\sigma v)d\Sigma_2^2,
\end{equation}
and this is the original form of the line element \cite{mdr86,mdr89}.
To transfer to non-null coordinates use $t\equiv av+br$,  requirement of a vanishing of
cross-terms gives $b=-a/(1+2\sigma)$ and line element
\begin{equation}\label{orignn}
ds^2=-\frac{(1+2\sigma)}{a^2}dt^2+\frac{dr^2}{(1+2\sigma)}
+r\left(\frac{r}{(1+2\sigma)}-\frac{2\sigma t}{a}\right)d\Sigma_2^2,
\end{equation}
where $a$ is an absorbable constant usually set to $1$.

One can transfer to potential coordinates where the scalar
field and Killing potential are the coordinates
\begin{eqnarray}\label{potentialcoords}
&ds^2=-\frac{1}{4y}dy^2+\frac{\sigma^2y}{{\rm sl}(x)^2}dx^2+\frac{\sigma^2y}{{\rm sl}(x)}d\Sigma_2^2,\\
&{\rm sl}(x)\equiv\sinh(x)(\sinh(x)+2\sigma\cosh(x)),~~~y=K,~~~x=\psi,~~~
\Omega=\left(1\pm\frac{\Lambda y}{12}\right)^{-1},\nonumber
\end{eqnarray}
also transferring to $t,z$ coordinates
\begin{equation}\label{tztrans}
t=\frac{1}{2\sigma}\ln(y),~
y=\exp(2\sigma t),~
z=\frac{1}{2\sigma}\ln\left(\frac{\sinh(x)^2}{{\rm sl}(x)}\right),~
\psi=x={\rm arccoth}(\frac{1}{2\sigma}(\exp(-2\sigma z)-1)),
\end{equation}
gives line element
\begin{equation}\label{tzcoords}
ds^2=\sigma^2\exp(2\sigma t)
\left[-dt^2+dz^2+\frac{(\exp(-2\sigma z)-1)^2-4\sigma^2}{4\sigma^2\exp(-2\sigma z)}d\Sigma_2^2\right].
\end{equation}

The hyperbolic dual solution is
\begin{eqnarray}\label{hyperbolicdual}
&ds^2=-\frac{\sigma^2y}{{\rm cl}(x)^2}dx^2+\frac{1}{4y}dy^2+\frac{\sigma^2y}{{\rm cl}(x)}d\Sigma_2^2,\\
&{\rm cl}(x)\equiv\cosh(x)(\cosh(x)+2\sigma\sinh(x))=1+{\rm sl}(x),
\Omega=\left(1\pm\frac{\Lambda y}{12}\right)^{-1},\nonumber
\end{eqnarray}
this just covers a region of the full solution where the Killing potential and scalar field
interchange being time and space coordinates.
Transferring to $t,z$ coordinates
\begin{equation}\label{dualcoordtrans}
z=\frac{1}{2\sigma}\ln(y),~
y=\exp(2\sigma z),~
t=\frac{1}{2\sigma}\ln\left(1+2\sigma\tanh(x)\right),~
\psi=x={\rm arctanh}(\frac{1}{2\sigma}(\exp(2\sigma t)-1)),
\end{equation}
gives line element
\begin{equation}\label{dualtz}
ds^2=\sigma^2\exp(2\sigma z)
\left[-dt^2+dz^2+\frac{(4\sigma^2-(\exp(2\sigma t)-1)^2)}{4\sigma^2\exp(2\sigma t)}d\Sigma_2^2\right].
\end{equation}
The solutions (\ref{potentialcoords}) and (\ref{hyperbolicdual}) correspond to different signs
of $K$ given by $u,v=+,-~or~-,+$,  the $K=0$ equivalently the $u~or~v=0$ part is not covered.
Note that defining
\begin{equation}\label{defm}
f\equiv\sqrt{\frac{1}{2}g_{\theta\theta}},~~~
m_a\equiv if4\delta_a^\theta+i\sin(\theta)f\delta_a^\phi,~~~
qm_{a b}\equiv m_a m^*_b+m^*_a m_b,~~~
ql_{a b}\equiv g_{a b}-qm_{a b},
\end{equation}
the Weyl tensor can be expressed as
\begin{equation}\label{weyl}
\frac{3}{R}C_{a b c d}
=g_{(a| c}g_{b| d)}
-3qm_{(a| c}qm_{b| d)}
-3ql_{(a| c}ql_{b| d)},
\end{equation}
in all cases,  so that there is no curvature indication of any difference between
(\ref{potentialcoords}) and (\ref{hyperbolicdual}).
For a possible causality difference between (\ref{potentialcoords}) and (\ref{hyperbolicdual})
note that $\psi_a\psi^a=g^{xx}$ so at first sight there might appear to be a difference,
however $y,{\rm sl},{\rm cl}$ can all change sign so it is not clear what happens.
A guess is that once a coordinate transformation has been found which turns
(\ref{hyperbolicdual}) into a form similar to (\ref{origsn}) it will turn out to
have self-similarity parameter $r/v$ instead of $v/r$.
\section{The Lanczos potential.}\lb{tlt}%%%%%%%%%%%%%%%%%%%%%%%%%%%%%%%%%%%%%%%%%%%%%%%%%%%%%%%
The field equations of general relativity can be re-written in a form analogous to
Maxwell's equations $F^{~b}_{a.;b}=J_a$ called Jordan's form \cite{HE} page (85)
\begin{equation}
C^{~~~d}_{abc.;d}=J_{abc},~~~
J_{abc}=R_{ca;b}-R_{cb;a}+\fr{1}{6}g_{cb}R_{;a}-\fr{1}{6}g_{ca}R_{;b},
\label{eq:1}
\end{equation}
where if field equations are assumed the Ricci tensor and scalar on the right hand side
can be replaced by stress tensors.
The Weyl tensor can be expressed in terms of the Lanczos potential \cite{Lanczos,mdrcs}
\begin{eqnarray}
\label{ldef}
\st{~}{C}_{abcd}&=&\st{1}{C}_{abcd}+\st{2}{C}_{abcd}+\st{3}{C}_{abcd}\\
\st{1}{C}_{abcd}&\equiv & H_{abc;d}-H_{abd;c}+H_{cda;b}-H_{cdb;a},~~
\st{3}{C}_{abcd}\equiv \frac{4}{(1-d)(2-d)}H^{ef}_{..e;f}(g_{ac}g_{bd}-g_{ad}g_{bc}),\nonumber\\
\st{2}{C}_{abcd}&\equiv & \frac{1}{(2-d)}\left\{g_{ac}(H_{bd}+H_{db})-g_{ad}(H_{bc}+H_{cb})+
    g_{bd}(H_{ac}+H_{ca})-g_{bc}(H_{ad}+H_{da})\right\},\nonumber
\end{eqnarray}
where the coefficients of $\st{2}{C}$ and $\st{3}{C}$ are fixed by requiring that the Weyl
tensor obeys the trace condition $C^{a}_{.bad}=0$.
Note that in \cite{mdr15} the $d=4$ values was assumed to hold in higher dimension.
$H_{bd}$ is defined by
\begin{equation}
H_{bd}\equiv H^{~e}_{b.d;e}-H^{~e}_{b.e;d}.
\label{eq:6}
\end{equation}
The Lanczos potential has the symmetries
\begin{equation}
2H_{[ab]c}\equiv H_{abc}+H_{bac}=0,~~~
6H_{[abc]}\equiv H_{abc}+H_{bca}+H_{cab}=0.
\label{eq:5}
\end{equation}
Equation (\ref{ldef}) is invariant under the algebraic gauge transformation
\begin{equation}
H_{abc}\rightarrow H'_{abc}=H_{abc}+\ch_a g_{bc}-\ch_b g_{ac},
\label{eq:7}
\end{equation}
where $\ch_a$ is an arbitrary four vector,
this transformation again fixes the coefficients of $\st{2}{C}$ and $\st{3}{C}$ .

In four dimensions the Lanczos potential with the above symmetries has twenty degrees
of freedom,  but the Weyl tensor has ten.
Lanczos \cite{Lanczos} reduced the degrees of freedom to ten
by choosing the algebraic gauge condition
\begin{equation}
3\ch_a=H^{~b}_{a.b}=0,
\label{lag}
\end{equation}
and the differential gauge condition
\begin{equation}
L_{ab}=H^{~~c}_{ab.;c}=0.
\label{ldg}
\end{equation}
The differential gauge transformation alters components which do not participate
in constructing the Weyl tensor:  contrast with electromagnetic theory where a gauge
transformation alters components in the vector potential all of which
participate in constructing the electromagnetic tensor.
The Lanczos potential has the weak field expansion \cite{Lanczos}
\begin{equation}
\st{weak}{H}_{abc}=\frac{1}{4}\left(h_{ac,b}-h_{bc,a}\right)
+\frac{1}{24}\left(h_{,a}\eta_{bc}-h_{,b}\eta_{ac}\right),
\label{lpweak}
\end{equation}
this expansion assumes that the current (\ref{eq:1}) vanishes $J=0$.

For line element (\ref{solution}) by trial and error a Lanczos potential is found to be
\begin{equation}
\label{ltdn}
H_{uvu}=\frac{u+3v+f}{6\Omega^3r_+r_-},~~~
H_{vuv}=\frac{3u+(1-4\sigma^2)v+f}{6\Omega^3r_+r_-},
\end{equation}
note that in this gauge the Lanczos potential has no angular term.
$f=0$ gives $\Omega^{-1}C$ to get the correct expression for the Weyl tensor it is necessary
to do a further integration to obtain $f$ the complex function
\begin{eqnarray}
\label{fequation}
f=&&\frac{1}{24\sigma^2}\left\{4u^3-9vu^2+6(1-4\sigma^2)v^2u-(1-4\sigma^2)^2v^3\right\}\\
&&+\frac{(-r_+r_-)^\frac{3}{2}}{16\sigma^3}\ln\left(2\sigma v +\sqrt{-r_+r_-}\right).\nonumber
\end{eqnarray}
%\section{Killing Spinor}\label{ks}%%%%%%%%%%%%%%%%%%%%%%%%%%%%%%%%%%%%%%%%%%%%%%%%%%%%%%%%%%%%
%\cite{BF}
%\cite{LPR}
%\cite{shuster}
%\cite{MB}
%Super symmetry can be expressed by
%\begin{equation}
%\label{supersymmetry}
%\delta\left(fermion\right)=\partial\left(boson\right)\epsilon,~~~
%\delta\left(boson\right)=\bar{\epsilon}\left(fermion\right),
%\end{equation}
%where $\epsilon$ is the spinor of transformation,  in the case of gravity it is a Killing
%spinor.
\section{Conclusion}\label{conc}%%%%%%%%%%%%%%%%%%%%%%%%%%%%%%%%%%%%%%%%%%%%%%%%%%%%%%%%%%%%%%%

In \S\ref{two} a two parameter solution (\ref{solution}) was found which has imploding spacetime 
and part of de Sitter spactime as special cases.  It demonstrates that the cosmological constant 
does not alter self-similarity.

In \S\ref{tps} a new three parameter scalar-Einstein solution (\ref{three}) was presented:
its properties are messy unlike those of the two parameter cases.

In \S\ref{hd} the hyperbolic dual (\ref{hyperbolicdual}) of the cosmological constant dynamical
scalar-Einstein  (\ref{solution}) was found,  it turns out that different hyperbolic coordinates
cover different parts of the solution depending on the signs of the scalar field and Killing 
potential.
Using a type of tetrad (\ref{defm}) the Weyl tensor (\ref{weyl}) takes the same form
in both cases.

In \S\ref{tlt} the Lanczos potential was found,  explicitly showing that the presence of the
conformal factor requires further integration.

Topics not looked at include the following.
The use of conformal factors to find further solutions \cite{mdr23}.
%The properties of the Lanczos potential \cite{mdrcs},
%which presumably are inherited from the $\Lambda=0$ case because of the vanishing
%of the Weyl tensor in deSitter spacetime.
Extension to higher dimensions,
this has been done for imploding scalar solutions \cite{mdrfi}.
Embedding in higher dimensions,  although this can be done in general for spherically symmetric
spacetimes a simple embedding for imploding scalar spacetime is not known.
The properties of geodesics:  deSitter spacetime is one of the few spacetimes in which all
geodesics can be easily expressed \cite{mdr20},  imploding scalar spacetime is exactly the
opposite not only is there no simple expression for the global geodesics but furthermore
local expressions for the world function and van Vleck determinant converge slowly.
The global properties of the solution (\ref{solution}),  although the global properties of the
limiting cases are known when the familiar coordinate transformations are applied one gets
off-diagonal metric components and so forth.
Group properties:  although the group properties of deSitter and anti-deSitter spectime are
well known nothing is known about the group properties of imploding scalar spacetime
or of (\ref{solution}).

In conclusion a solution which has both imploding scalar and parts of anti-deSitter or deSitter
spacetime as limiting cases was presented and its null junction conditions shown to be well behaved.
%$\Lambda=0$ solution was shown to have
%have good junction conditions with Minkowski spacetime but for $\Lambda\ne0$ this seems to
%no longer be the case.


\begin{thebibliography}{99}%%%%%%%%%%%%%%%%%%%%%%%%%%%%%%%%%%%%%%%%%%%%%%%%%%%%%%%%%%%%%%%%%%%%



\bibitem{AKL}
M.A. Abramowicz, W. Klu\'ziak \& J.-P. Lasota,
No observational proof of the black-hole event-horizon.
{\it A\&A}{\bf 396}(2002)L31-L34.

\bibitem{BNS}
R.Baier,  H. Nishimura, S.A Stricker,
Scalar field collapse with negative cosmological constant.
\href{https://arxiv.org/abs/1410.3495}
                       {\tt 1410.3495}


\bibitem{BBGB}
Ivan S.N. Booth,  Lionel Brits, Jose A. Gonzalez,  Chris Van Den Broeck,
Marginally trapped tubes and dynamical horizons,
\href{https://arxiv.org/abs/gr-qc/0506119}
                       {\tt gr-qc/0506119}

\bibitem{brady}
Patrick R. Brady,
Analytic examples of critical behaviour in scalar field collapse.
{\it Class.Q.Grav.}{\bf 11}(1994)1255.

%\bibitem{BF}
%P. Breitlohner and D.Z. Freedman,
%Stabikity in guaged extended supergravity,
%{\it Ann.Phys.}{\bf 144}(1982)249-281.
%\href{https://doi.org/10.1016/0003-4916(82)90116-6}{doi}.

\bibitem{chase}
J.E. Chase,
Event horizons in static scalar-vacuum space-times,
{\it Commun.Math.Phys.}{\bf 19}(1970)276-288.

\bibitem{bi:chr}
Demetrios L. Christodoulou,
A mathematical theory of gravitational collapse,
{\it Comm.Math.Phys.},{\bf 109}(1987)614-647.

\bibitem{clifton}
Timothy Clifton,  David F. Mota,  John D. Barrow,
Inhomogeneous Gravity.
{\it Mon. Not. R. Astron Soc.}(2004)1-15,
\href{https://arxiv.org/abs/gr-qc/0406001}
                       {\tt gr-qc/0406001}

\bibitem{CS}
Toby Crisford and Jorge E. Santos,
Violating weak cosmic censorship in $ADS_4$,
\href{https://arxiv.org/abs/1702.05490}
                       {\tt 1702.05490}

\bibitem{frolov}
Andrei V. Frolov,
Critical Collapse Beyond Spherical Symmetry: General Perturbations of the Roberts Solution.
{\it Phys.Rev.D}{\bf 59}(1999)104011-1-7.
\href{https://arxiv.org/abs/gr-qc/9811001}
                       {\tt gr-qc/9811001}

\bibitem{GJ}
Alexander A.H. Graham and Rahul Jha,
Stationary black holes with time-dependent scalar fields..
\href{http://arXiv.org/abs/1407.6573}
                      {\tt 1407.6573}

\bibitem{gundlach}
Carsten Gundlach,
Critical Phenomena in Gravitational Collapse.
\href{http://arXiv.org/abs/gr-qc/9712084}
                      {\tt gr-qc/9712084}
{\it Adv.Theor.Math.Phys.}{\bf 2}(1998)1-49.

\bibitem{HE}
S.W. Hawking and G.F.R. Ellis,
The Large Scale Structure of Space-Time,
Math.Rev.\href{http://www.ams.org/mathscinet-getitem?mr=54:12154}
                                                   {\sl 54 \#12154},
Cambridge University Press,  Cambridge (1973).

\bibitem{Lanczos}
Cornelius Lanczos,
The splitting of the Riemann tensor,
{\it Rev.Mod.Phys.}{\bf 34}(1962)379.

%\bibitem{LPR}
%H. L\"u,  C.N. Pope and J. Rahmfeld,
%A construction of Killing spinors on %S^n$,
%\href{https://arxiv.org/abs/gr-qc/9805151}
%                       {\tt gr-qc/9805151}

%\bibitem{MB}
%R.G. McLenaghan and N. Van den Bergh,
%Spacetimes admitting Killing 2-spinors,
%{\it Class.Q.Grav.}{\bf 10}(1993)2179-2185

\bibitem{NM}
Ramesh Narayan and Jeffrey E. McClintock,
Observational Evidence for Black Holes.
\href{http://arXiv.org/abs/1312.6698}
                      {\tt 1312.6698}

\bibitem{choptuik}
David W. Neilsen A and Matthew W. Choptuik,
Critical phenomena in perfect fluids.
\href{http://arXiv.org/abs/gr-qc/9812053}
                      {\tt gr-qc/9812053}

\bibitem{mdr86}
Mark D. Roberts,
PhD Thesis,
Spherically symmetric fields in gravitational theory.
University of London (1986).

\bibitem{mdr89}
Mark D. Roberts,
Scalar Field Counter-Examples to the Cosmic Censorship Hypothesis.
{\it Gen.Rel.Grav.}{\bf 21}(1989)907-939.

\bibitem{mdr15}
Mark D. Roberts,
Dimensional Reduction and the Lanczos Tensor.
{\it Mod.Phys.Lett.}
{\bf A4}(1989)2739-2746.

\bibitem{mdr20}
Mark D. Roberts,
The World Function in Robertson-Walker Spacetime.
{\it Astrophys.Lett. \& Commun.}
{\bf 28}(1993)349-357.

\bibitem{mdr23}
Mark D. Roberts,
Imploding Scalar Fields.
{\it J.Math.Phys.}
{\bf 37}(1996)4557-4573.

\bibitem{mdrfi}
Mark D. Roberts,
Interacting with the Fifth Dimension.
{\it Phys.Astron.Int.J.}2(3)(2018):00054,
\href{http://arXiv.org/abs/gr-qc/0412051}
                      {\tt gr-qc/0412051}

\bibitem{mdrfd}
Mark D. Roberts,
Is spacetime non-metric?
{\it Expert Opinion on Astronomy and Astrophysics}, To appear,
\href{http://arXiv.org/abs/0706.4043}
                      {\tt 0706.4043}

\bibitem{mdrcs}
Mark D. Roberts,
The Lanczos potential and Chern-Simons theory.
\href{http://arXiv.org/abs/0808.1687}
                      {\tt 0808.1687}

\bibitem{mdrqs}
Mark D. Roberts,
Quantum Imploding Scalar Fields.
\href{http://rsos.royalsocietypublishing.org/content/5/10/180692}
{\it R. Soc. open sci.},{\bf 5}(2018)5:180692,
\href{http://dx.doi.org/10.1098/rsos.180692}{\color{yellow}doi}
\href{https://arxiv.org/abs/1703.04579}
                       {\tt 1703.04579}

%\bibitem{shuster}%%%no journal
%Eugene Shuster,
%Killing spinor and supersymmetry on ads,
%\href{https://arxiv.org/abs/hep-th/9902120}
%                       {\tt hep-th/9902129}

\bibitem{szabodos}%23.7
Laszlo B. Szabodos,
Math.Rev.\href{http://www.ams.org/mathscinet-getitem?mr=90h:83066}
{\sl 90h:83066}.

\bibitem{torres}
Observations of superradiance in a vortex flow.
Theo Torrs, Sam Patrick, Antonin Coutant, Maurice Richartz, Edmund W. Tedford, Silke Weinfurtner,
{\it Nature}{\bf 13}(2017)833-836,
\href{https://arxiv.org/abs/1612.06180}
                       {\tt 1612.06180}


\end{thebibliography}
\end{document}